\UseRawInputEncoding
\documentclass[prb,reprint,amsmath,amssymb,aps,twocolumn,superscriptaddress]{revtex4-1}
\usepackage{hyperref}
\usepackage{rotating}
\usepackage{graphicx}
\usepackage{latexsym}
\usepackage{amssymb}
\usepackage{amsfonts}
\usepackage{soul}
\usepackage{color}
\usepackage{amsmath}
\usepackage{color}
\usepackage{lipsum}
\usepackage{natbib}

\usepackage{changes}

\begin{document}

\title{Suppression of fluctuations in a two-band superconductor with a quasi-1D band}

\author{A. A. Shanenko}
\email{ashanenko@hse.ru}
\affiliation{HSE University, 101000, Moscow, Russia}

\author{T. T. Saraiva}
\affiliation{HSE University, 101000, Moscow, Russia}

\author{A. Vagov}
\affiliation{HSE University, 101000, Moscow, Russia}
\affiliation{Institut f\"{u}r Theoretische Physik III, Bayreuth Universit\"{a}t, Bayreuth 95440, Germany}

\author{A. S. Vasenko}
\affiliation{HSE University, 101000, Moscow, Russia}

\author{A. Perali}
\affiliation{School of Pharmacy, Physics Unit, University of Camerino, I-62032 Camerino, Italy}

\date{\today}
\begin{abstract}
Chain-like structured superconductive materials (such as A$_2$Cr$_3$As$_3$, with A = K, Rb, Cs) exhibit the multiband electronic structure of single-particle states, where coexisting quasi-one-dimensional (Q1D) and conventional higher-dimensional energy bands take part in the creation of the aggregate superconducting condensate. When the chemical potential approaches the edge of a Q1D band in a single-band superconductor, the corresponding mean-field critical temperature increases significantly but the superconductivity is quenched by fluctuations. However, recent investigation has revealed that when a Q1D band is coupled to a higher dimensional one by the interband Cooper-pair transfer, the thermal superconductive fluctuations can be suppressed so that the resulting critical temperature can be close to its mean-field value. In the present work, we calculate the mean-field $T_{c0}$ and fluctuation-shifted $T_c$ critical temperatures for a two-band superconductor where a Q1D band coexists with a higher-dimensional band, and investigate how the thermal fluctuations are sensitive to the system parameters. We find that $T_{c}$ is close to $T_{c0}$ in a wide range of microscopic parameters, and even the dimensionality of the higher-dimensional band does not play an essential role. Thus, the screening mechanism for suppressing fluctuations via the pair-exchange coupling between the bands is indeed relevant for a large class of Q1D multiband superconducting materials, encouraging further experiments aimed at reaching larger critical temperatures in such multiband superconductors. 
\end{abstract}

\maketitle
\section{Introduction}

Experimental detection of two superconductive gaps in MgB$_2$~\cite{Bouquet2001a,Bouquet2001b, Karapetrov2001, Iavarone2002} sparked a plethora of theoretical studies of the multiband or multi-gap models of superconductivity, see e.g. Refs.~\onlinecite{Liu2001, Askerzade2001, Golubov2002, Mazin2003, Konsin2004, Askerzade2006}. The core of the difference between the multi- and single-band materials lies in the interference of multiple contributing condensates, which makes their properties to deviate from those of the single-condensate systems. The interference can, among other things, suppress superconducting order-parameter fluctuations, which is here referred to as the multiband fluctuations screening mechanism~\cite{Salasnich2019, Saraiva2020}. In particular, recently it has been demonstrated~\cite{Saraiva2020} that enormous thermal fluctuations in a Q1D superconducting condensate \cite{Efetov1974,Gorkov1975,Klemm1976} can be suppressed almost completely when the latter is coupled to a 3D condensate via the pair-exchange transfer, even when this coupling is rather weak~\cite{Saraiva2020}. 

The study in Ref.~\onlinecite{Saraiva2020} has also demonstrated that the fluctuation screening can have a significant effect even in the case of a shallow Q1D band coupled to a conventional deep 3D band (shallow and deep refers to a position of the chemical potential, close or far from the band edge, respectively). The superconducting condensate in such a case is a coherent mixture of the standard BCS state in the 3D deep band and nearly BEC state in the Q1D shallow band -- the so-called multiband BCS-BEC crossover regime, characterized by a much higher mean-field critical temperature. The physical reason for this amplification of the superconducting temperature is a Feshbach-like resonance, see e.g. Refs.~\onlinecite{Perali1996, Bianconi1997, Shanenko2006, Shanenko2010, Innocenti2010, Bianconi2013, Mazziotti2017}. It appears when the chemical potential approaches the Q1D-band edge, which results in much higher density of single-particle states (DOS) due to the van Hove singularity. The fluctuations of the superconducting order parameter in Q1D band tend to quench the superconductivity~\cite{Efetov1974, Gorkov1975, Klemm1976} and their impact further increases in the presence of a shallow Q1D band. However, the fluctuation screening induced by the pair-exchange coupling to a band of a larger dimensionality can suppresses the fluctuations and restore the superconductivity, also at elevated temperatures $T\sim T_{c0}$. This scenario to reach a high critical temperature is of especial relevance for materials that combine Q1D and higher-dimensional bands, like the recent chain-like structured materials~\cite{Jiang2015, Wu2019,Xu2020}. However, the same screening mechanism takes place also in systems where both shallow and deep bands are quasi-2D (Q2D) \cite{Salasnich2019}. 

Motivated by recent theoretical works reporting this fluctuation suppression mechanism~\cite{Salasnich2019, Saraiva2020} and ongoing experiments on the multiband superconductors with Q1D bands, such as A$_2$Cr$_3$As$_3$ (A = K, Rb, Cs) \cite{Jiang2015,Bao2015,Zhi2015,Tang2015A,Tang2015B,Wu2019,Xu2020} and organic superconducting compounds \cite{Liu2017, Haoxiang2019,Yan2019,Pinto2020}, we investigate details of the fluctuation screening in a two-band system, where a shallow Q1D band is coupled to a conventional Q2D or 3D reservoir band. This model serves as a prototype for the chain-like structured superconducting materials mentioned above. 

This work complements the earlier study~\cite{Saraiva2020} by investigating how the fluctuation-induced renormalization of the critical temperature depends on the interplay of the microscopic parameters such as the dimensionality of the higher-dimensional reservoir band, its energy depth (the Fermi energy) and intraband interaction. These parameters determine the fluctuations suppression in the Q1D band by controlling the fluctuations in the reservoir band. The latter become stronger when e.g. the energy depth or the dimensionality of the reservoir band decrease. It is thus necessary to go beyond a simplified model considered in the previous study~\cite{Saraiva2020} where only the coupling of the Q1D condensate to that of the 3D deep band (with almost negligible fluctuations) has been investigated.

The paper is organized as follows. In Sec. \ref{sec2} we consider the two-band generalization of the BCS model and the equation for the mean-field critical temperature. Then, we derive the effective Ginzburg-Landau (GL) free energy functional that controls the superconducting order-parameter fluctuation corrections to the critical temperature. Details of the computing the Q1D coefficients in the GL functional are given in Appendix \ref{secApp}. In Sec.~\ref{sec3} we discuss relevant parameters of the two-band system and calculate the mean-field critical temperature $T_{c0}$ and its fluctuation-renormalized value $T_c$. Section \ref{sec4} summarizes our results. 

\section{Formalism}
\label{sec2}

In this section, we outline the formalism necessary to calculate the mean-field $T_{c0}$ and fluctuation-renormalized $T_c$ critical temperatures of a two-band superconductor. We assume that one of the bands is quasi-one-dimensional (Q1D) and it is close to the Feshbach-like resonance associated with the Lifshitz transition~\cite{Perali1996, Bianconi1997, Shanenko2006, Shanenko2010, Innocenti2010, Bianconi2013, Mazziotti2017} that occurs when the chemical potential crosses the bottom of the Q1D band. This band can be referred to as shallow. The second band has a higher dimensionality and its energy depth (the Fermi energy) is varied in our calculations. We consider two variants of this band - quasi-two-dimensional (Q2D) and three dimensional (3D). The Q1D band is assumed to determine the mean-field critical temperature $T_{c0}$ of the system. The intraband coupling in the higher-dimensional band is weak enough, so that the critical temperature of the superconductive transition in this band, taken as a separate superconductor, is much lower than $T_{c0}$ of the two-band system. 

\subsection{Two-band BCS model}

We consider the two-band system with the $s$-wave pairing in both bands, using the standard generalization of the BCS model\cite{Suhl1959,Moskalenko1959} with the pair-exchange coupling between the different bands. The coupling matrix $g_{\nu\nu'}$~($\nu,\nu'=1,2$) is symmetric and real, where $\nu=1$ stands for the higher-dimensional (Q2D or 3D) band and $\nu=2$ corresponds to the Q1D band. We consider that the system is in the clean limit and the effects of impurities can be neglected. The mean-field Hamiltonian of the model in the real space writes as~\cite{Shanenko2011}
\begin{align}
\mathcal{H}=&\int d^3{\bf r}\Bigg\{\sum_{\nu=1,2} \Bigg[\sum_{\sigma=\uparrow,\downarrow}\hat{\psi}_{\nu\sigma}^\dagger({\bf r})T_\nu({\bf r})\hat{\psi}_{\nu\sigma}({\bf r})\nonumber\\
&+\left(\hat{\psi}_{\nu\uparrow}^\dagger({\bf r})\hat{\psi}_{\nu\downarrow}^\dagger({\bf r}) \Delta_\nu({\bf r}) + {\rm h.c.}\right)\Bigg] + \langle \vec\Delta, \breve{\gamma}\,\vec\Delta \rangle \Bigg\},
\label{eq.H}
\end{align}
where  ${\hat \psi}_{\nu \sigma}({\bf r})$ are the operators for carriers with spin $\sigma $ in band $\nu$, $T_{\nu}({\bf r})$ is the single-particle Hamiltonian, and $\Delta_{\nu}({\bf r})$ is the gap function in the respective band. Here we use the vector notation $\vec \Delta = (\Delta_1,\Delta_2)^T$, with the scalar product $\langle.,.\rangle$, and denote by $\breve{\gamma} = \breve{g}^{-1}$ the inverse of the coupling matrix $\breve{g}$ with elements $g_{\nu\nu'}$. 

The mean-field solution for the condensate is obtained by diagonalizing the Hamiltonian self-consistently, together with the self-consistency gap equation \cite{Shanenko2011,Vagov2012B} 
\begin{equation}
\Delta_\nu({\bf r})=\sum_{\nu'=1,2}g_{\nu\nu'}R_{\nu'}({\bf r}),\label{eq.selfcons}
\end{equation}
where $R_\nu({\bf r}) = \langle\psi_{\nu\uparrow}(\bf r)\psi_{\nu \downarrow}(\bf r)\rangle$ are the anomalous averages. 

For the single-particle Hamiltonian $T_{\nu}({\bf r})$ we adopt the effective mass approximation \cite{Saraiva2020} so that the single-particle energy is approximated as
\begin{equation}
\xi_{{\bf k}1}=\varepsilon_0+\sum\limits_\alpha \frac{\hbar^2 k_\alpha^2}{2m_1}-\mu, \;\,\xi_{{\bf k}2}=\frac{\hbar^2 k_x^2}{2m_2}-\mu,
\label{eq.xinu}
\end{equation}
where ${\bf k}=\{k_x,k_y,k_z\}$, and index $\alpha$ assumes values $\{x,y\}$ for the Q2D band and $\{x,y,z\}$ for the 3D one; $m_{\nu}$ is the carrier effective mass for band $\nu$, $\varepsilon_0$ is the energy shift between the bands, and $\mu$ is the chemical potential of the system, measured with respect to the edge of the Q1D band. Notice that the energy dispersion of the Q1D (Q2D) band is degenerate in the remaining $\{y,z\}$ ($z$) directions. The respective summation over $k_\alpha$ in these directions, which is needed in the calculations of the band density of states (DOS), is estimated by simply introducing multiplicative factors $n_y$ and $n_z$ related to the Brillouin zone sizes in the corresponding directions. 

\begin{figure}[t]
\centering
\includegraphics[width=\linewidth]{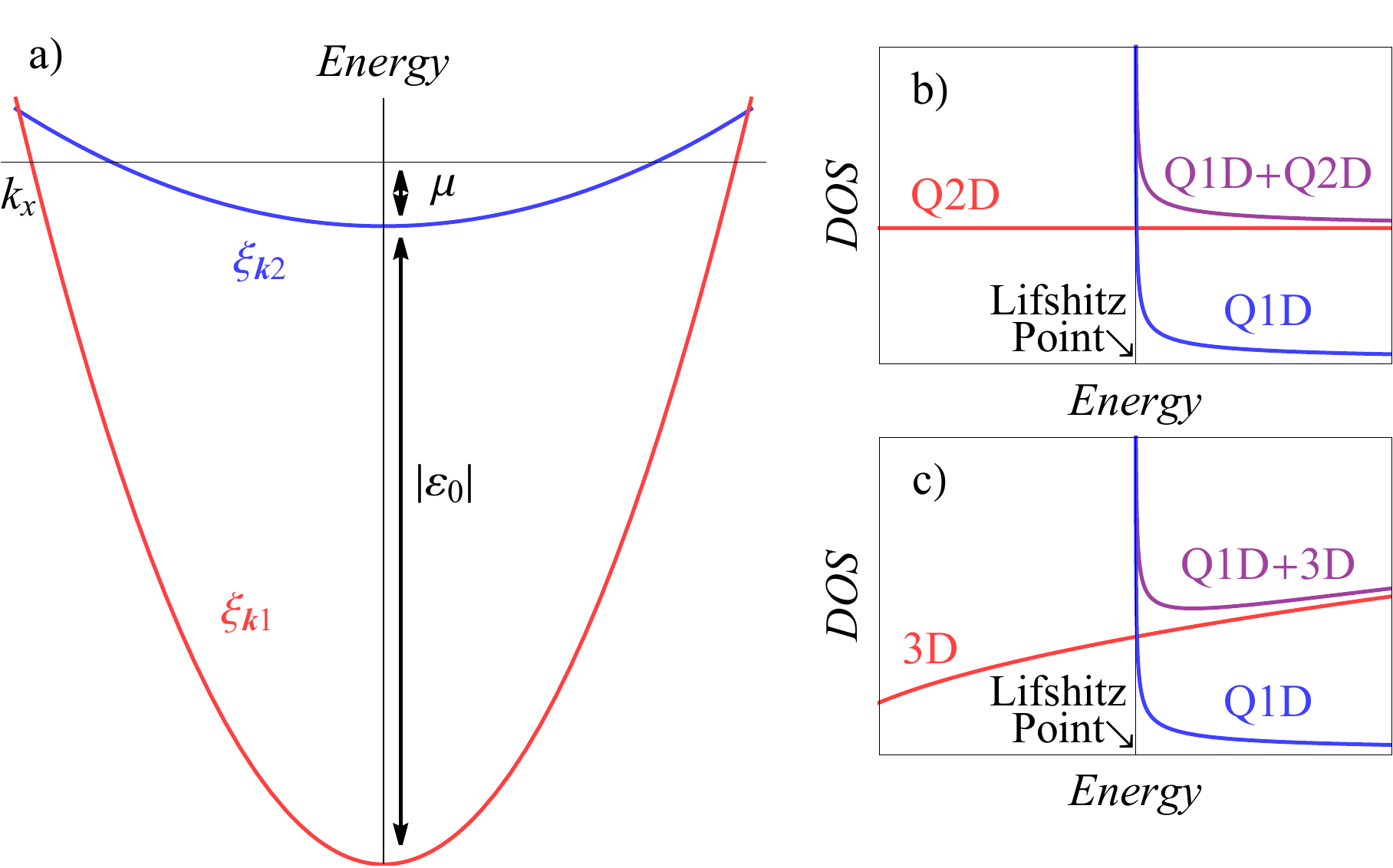}
\caption{\label{fig1} a) Sketch of the single-particle energies $\xi_{{\bf k} 1}$ and $\xi_{{\bf k} 2}$ versus $k_x$. Panels b) and c) illustrate the energy-dependent DOSs for the Q1D+Q2D and Q1D+3D two-band systems: the total and band-dependent DOSs are shown in a range of energies around the Lifshitz point of the order of the cut-off energy.}
\end{figure}

The energies and the chemical potential are taken relative to the edge of the Q1D band. The edge of the higher-dimensional band is located below the edge of the Q1D band, so that $\varepsilon_0 < 0$. A schematic energy diagram of the bands is shown in Fig.~\ref{fig1} a). The calculations involving the higher-dimensional band are performed within the standard approximations of the BCS theory, whereas the contribution of the shallow band requires a more accurate approach. In what follows we assume $|\mu| < \hbar\omega_c$, with $\hbar\omega_c$ the cut-off energy of the pairing interaction, where the Q1D Feshbach-like resonance is most pronounced. Below we set the Boltzmann constant as $k_B=1$.

\subsection{Mean field critical temperature}

The mean field critical temperature $T_{c0}$ is obtained by solving the linearized gap equation \begin{equation}
\sum_{\nu'=1,2}L_{\nu\nu'}\Delta_{\nu'}=0,\; L_{\nu\nu'}=\gamma_{\nu\nu'}-\mathcal{A}_\nu\delta_{\nu\nu'},
\label{eq.gapex}
\end{equation}
where $\delta_{\nu\nu'}$ is the Kronecker symbol, and $\mathcal{A}_{\nu}$ are given by (see Appendix~\ref{secApp})
\begin{align}
\mathcal{A}_{1}=N_1\ln\left(\frac{3.56}{\pi  T_{c0}}\right), \,
\mathcal{A}_2=N_2\int\limits_{- \tilde\mu}^{1}d\varepsilon\frac{\tanh( \varepsilon/2 \tilde T_{c0} )}{\varepsilon \sqrt{ \varepsilon + \tilde\mu}}
\label{eq.dos},
\end{align}
where the quantities marked by a tilde are normalized by the cut-off energy, the DOS $N_1$ for the higher-dimensional band is given by $N_1^{Q2D}= n_z m_1/2\hbar^2$ for the Q2D case and $N_1^{3D}=m_1 k_F/2\pi^2\hbar^2$ for the 3D case, with the Fermi momentum $\hbar k_F =\sqrt{2m_1(\mu+|\varepsilon_0|)}$. For the Q1D band $N_2= n_y n_z \sqrt{m_2/32\pi^2 \hbar^3 \omega_c}$, which is the Q1D DOS at the cut-off energy $\hbar\omega_c$~(the divergent part of the energy-dependent Q1D DOS is kept inside the integral). The energy-dependent DOSs for bands $1$ and $2$ are sketched in Figs.~\ref{fig1}. For simplicity and without loss of generality, we assume that the factor $n_z$ is the same for both bands. 

The critical temperature $T_{c0}$ is found from Eq.~(\ref{eq.gapex}). The existence of a nontrivial solution for the gap functions assumes that the determinant of the matrix $\breve{L}$~(with the elements $L_{\nu\nu'}$) is zero, and one gets
\begin{equation}
\left(g_{22}-G\mathcal{A}_1\right)\left(g_{11}-G\mathcal{A}_2\right)-g_{12}^2=0,
\label{eq.Tc0}
\end{equation}
where $G=g_{11}g_{22}-g_{12}^2$. Since ${\cal A}_\nu \propto N_\nu$, the solution depends on the dimensionless coupling constants
\begin{equation}
\lambda_{11}=g_{11}N_1,\;
\lambda_{22}=g_{22}N_2,\;
\lambda_{12}=g_{12}\sqrt{N_1N_2}.
\label{eq.lambda}
\end{equation}
Of the two possible solutions to Eq. (\ref{eq.Tc0}) one has to choose the one with the largest $T_{c0}$. Notice that the choice of $m_1$, $m_2$, $n_y$ and $n_z$ is not important here, one needs only to choose the dimensionless coupling constants $\lambda_{ij}$ to calculate $T_{c0}$. 

\subsection{The free energy functional}

The mean-field results of the previous section can be strongly modified by thermal superconducting fluctuations. To investigate the impact of those fluctuations by calculating the related corrections to the critical temperature in the vicinity of the Lifshitz transition $\mu \simeq 0$, we evaluate the corrections by using the Gibbs distribution $e^{-F/T}$ with the free energy $F$ given by
\begin{equation}
F = \int d^3{\bf r}\Big[\sum_{\nu=1,2} f_\nu  +  \langle\vec\Delta, \check{L}
\vec\Delta\rangle\Big],    
\label{eq.free1}
\end{equation}
where $f_{\nu}$ in the vicinity of $T_{c0}$ can be expanded in powers of $\Delta_{\nu}$ and its gradients, which yields [see Appendix \ref{secApp}] 
\begin{equation}
f_\nu = a_\nu \left|\Delta_{\nu} \right|^2  + \frac{b_{\nu}}{2} \left|
\Delta_{\nu} \right|^4 +\sum\limits_{\alpha=x,y,z} {\cal K}^{(\alpha)}_{\nu}  \left|\partial_\alpha \Delta_{\nu}
\right|^2.
\label{eq.free2}
\end{equation}
For the higher-dimensional band the coefficients are given by the standard expressions. In particular, $a_1$ and $b_1$ are the same for the Q2D and 3D variants  
\begin{align}
a_1 = -\tau N_1,\quad b_1 = \frac{7\zeta(3)}{8\pi^2}\frac{N_1}{T^2_{c0}}
\label{eq.coeff1-ab}
\end{align}
with $\tau=1-T/T_{c0}$. The remaining coefficient is given by
\begin{align}
{\cal K}^{(\alpha)}_1=\frac{\hbar^2v_1^2}{6} b_1
\label{eq.coeff1-K3D}
\end{align}
with $\alpha = \{x,y,z\}$ for the 3D band and
\begin{align}
{\cal K}^{(\alpha)}_1 = \frac{\hbar^2v_1^2}{4} b_1,\quad
{\cal K}^{(z)}_1=0,
\label{eq.coeff1-K2D}
\end{align}
with $\alpha = \{x,y\}$ for the Q2D band. For the Fermi velocity in these expressions we have $v_1 = \hbar k_F/m_1$. 

For the shallow Q1D band (for $|\mu| < \hbar\omega_c$) the expressions for the coefficients can only be represented in the form of the integrals written as
\begin{align}
& a_2 = -\tau\frac{N_2}{2 T_{c0}} \int\limits_{- \tilde{\mu}}^1 d\varepsilon \, \frac{\text{sech}^2  \big( \varepsilon / 2\tilde{T}_{c0} \big) }{\sqrt{\varepsilon+ \tilde{\mu}}}, \notag \\
& b_2 =  \frac{N_2}{4 \hbar^2 \omega_c^2} \int \limits_{- \tilde{\mu}}^1 d\varepsilon\,  \frac{\text{sech}^2  \big(\varepsilon/2  \tilde{T}_{c0} \big) }{\varepsilon^3 \sqrt{\varepsilon + \tilde{\mu}}}  \left[ \sinh \Big( \frac{\varepsilon}{ \tilde{T}_{c0}} \Big)- \frac{\varepsilon}{ \tilde{T}_{c0}}\right], \notag \\
& \mathcal{K}^{(x)}_2 =  \frac{N_2 v_2^2}{8\,  \omega_c^2} \int\limits_{- \tilde{\mu}}^1 d\varepsilon\, \frac{\sqrt{\varepsilon + \tilde{\mu}} }{\varepsilon^3} \, \text{sech}^2  \left(\frac{\varepsilon}{2\tilde{T}_{c0}}\right)  \notag \\
& \quad \quad \times \left[ \sinh \left( \frac{\varepsilon}{ \tilde{T}_{c0}} \right) - \frac{\varepsilon}{\tilde{T}_{c0}}\right], \quad  {\cal K}^{(y,z)}_2=0,
\label{eq.coeff2}
\end{align}
where $\tilde{T}_{c0}$ and $\tilde{\mu}$ are defined in Eq.~(\ref{eq.dos}), and the characteristic velocity of the Q1D band is given by $v_2 = \sqrt{2 \hbar \omega_c/m_2}$.

The free energy in Eq.~(\ref{eq.free1}) for the two-band system can be simplified considerably by representing $\vec{\Delta}$ as a linear combination of the eigenvectors of the matrix $\breve{L}$ as \cite{Salasnich2019,Saraiva2020}
\begin{equation}
\vec{\eta}_+=
\left(\begin{array}{c}
S\\
1
\end{array}\right),\;\vec{\eta}_-=
\left(\begin{array}{c}
1\\
-S
\end{array}\right),
\label{eq.eta}
\end{equation}
where
\begin{align}
S=\frac{g_{11}-G\mathcal{A}_2}{g_{12}},
\label{eq.S}
\end{align}
($S \geq 0$ for the $s$-wave pairing). Using the representation
\begin{align}
\vec{\Delta}({\bf r}) = \psi ({\bf r})\vec{\eta}_+ +\varphi({\bf r}) \vec{\eta}_-,
\label{eq.exp}
\end{align}
where $\psi({\bf r})$ and $\varphi({\bf r})$ are the modes associated with $\vec{\eta}_+$ and $\vec{\eta}_-$, the free energy functional can be rearranged as 
\begin{align}
F  = \int d^3{\bf r}(f_\psi  + f_\varphi + f_{\psi \varphi}),
\label{eq.free3}
\end{align}
where $f_\psi$ and $f_\varphi$ have the same structure as Eq.~(\ref{eq.free2}), but with $\Delta_\nu$ replaced by $\psi({\bf r})$ and $\varphi({\bf r})$, respectively. The coefficients in $a_\nu, b_\nu, {\cal K}^{(\alpha)}_{\nu}$ are changed as
\begin{align}
&a_{\psi}=S^2  a_1 + a_2,\; b_{\psi} = S^4 b_1 + b_2, \notag\\
&{\cal K}^{(\alpha)}_{\psi} = S^2{\cal K}^{(\alpha)}_1 + {\cal K}^{(\alpha)}_2
\label{eq.coeff-psi}
\end{align}
and  
\begin{align}
&a_{\varphi} = a^{(0)}_{\varphi}+ a_1 +S^2 a_2, \; b_{\varphi} = b_1 + S^4b_2,\notag\\
&{\cal K}^{(\alpha )}_{\varphi} = {\cal K}^{(\alpha)}_1 + S^2{\cal K}^{(\alpha)}_2, \quad a^{(0)}_{\varphi} = \frac{(1+S^2)^2}{S G g_{12}},
\label{eq.coeff-phi}
\end{align}
with $\alpha = \{x,y,z\}$. Finally, $f_{\psi\varphi}$ in Eq.~(\ref{eq.free3}) describes the interaction between the modes $\psi$ and $\phi$. 

By virtue of Eq.~(\ref{eq.S}), the quantity $S$ is real and we have $a^{(0)}_{\varphi}\not = 0$ for arbitrary parameters of the two-band model. This implies that the characteristic length $\xi^{(\alpha)}_{\phi}=\sqrt{{\cal K}^{(\alpha)}_{\phi}/a_{\phi}}$ of the mode $\phi$ is generally finite near $T_{c0}$, which is a consequence of the fact that the two contributing condensates are coupled by the Josephson-like pair transfer between the band condensates. Consequently, $\psi$ is the only critical mode with the divergent characteristic length $\xi^{(\alpha)}_{\psi}=\sqrt{{\cal K}^{(\alpha)}_{\psi}/a_{\psi}}$ at $T \to T_{c0}$. The pair fluctuations, controlled by the mode $\varphi$, produce noncritical corrections, which can be safely neglected close to $T_{c0}$. Thus, the analysis of the pair fluctuations can consider only the critical mode, i.e. $F$ is well approximated by the single-component GL functional 
\begin{equation}
F \simeq \int d^3{\bf r}\Big(a_\psi |\psi|^2  + \frac{b_{\psi}}{2}|\psi|^4 +\sum\limits_{i=x,y,z} {\cal K}^{(\alpha)}_{\psi}|\partial_\alpha \psi|^2\Big),  
\label{eq.free4}
\end{equation}
where the presence of the two bands is reflected only the coefficients $a_\psi,b_\psi$, and ${\cal K}^{(\alpha)}_{\psi}$ that are averages over the contributing bands, see Eq.~(\ref{eq.coeff-psi}). 

From the definition of $\vec{\eta}_+$ in Eq.~(\ref{eq.eta}) it follows that $S$ controls the relative occupation of the reservoir band with $\nu=1$. In the limit $S \to \infty$, the Q1D band ($\nu = 2$) is depleted and $a_\psi \to  S^2 a_1$, $b_\psi \to  S^4 b_1$, and ${\cal K}^{(\alpha)}_\psi \to S^2 {\cal K}^{(\alpha)}_1$. Also, Eq.~(\ref{eq.eta}) yields $\Delta_1 = \psi S$ in this case, and the GL free energy of the system is reduced to the free energy of the higher-dimensional band. 
 
In the opposite limit $S \to 0$, the higher-dimensional band does not contribute. In this case, Eq. (\ref{eq.coeff-psi}) yields $a_\psi \to  a_2$, $b_\psi \to  b_2$, and ${\cal K}^{(\alpha)}_\psi \to {\cal K}^{(\alpha)}_2$. Now Eq.~(\ref{eq.eta}) gives $\psi = \Delta_2$, and the GL free energy in Eq.~(\ref{eq.free4}) is fully determined by the Q1D band. 

Before proceeding further, it is important to discuss Eq.~(\ref{eq.free4}) in the context of the fluctuation screening mechanism~\cite{Salasnich2019,Saraiva2020}. One sees that the thermal fluctuations in both bands are not independent as they are controlled by the same mode $\psi$: the band gap functions are given by $\delta\Delta_1=S \delta\psi$ and $\delta\Delta_2= \delta\psi$. Loosely speaking, ``light" fluctuations of the Q1D condensate are ``pinned"  to the ``heavy" fluctuations in the higher-dimensional band. This illustrates the physical reason why the fluctuations in the Q1D band can be screened by the reservoir higher-dimensional band. 

In more detail, the fluctuations are controlled by the superfluid stiffness coefficient ${\cal K}^{}_\psi$, defined by Eq.~(\ref{eq.coeff-psi}) as the average over the bands. One can see that in the limit $v_1 \gg  v_2$, the main contribution to the stiffness coefficient is provided by the higher-dimensional band. In this limit strong fluctuations, specific to Q1D systems, are fully suppressed and cannot affect the critical temperature~\cite{Saraiva2020}. However, in real systems the ratio $v_1/v_2$ is finite and the fluctuations reduce the critical temperature $T_c < T_{c0}$. Thus, the main problem is to clarify the domain of microscopic parameters of the model, where $T_c$ is not significantly reduced with respect to $T_{c0}$. 

\subsection{Ginzburg number}
\label{subs:Ginznum}

The impact of the thermal fluctuations on the critical temperature is determined by the Ginzburg number (also known as the Ginzburg-Levanyuk parameter) $Gi=1-T_{Gi}/T_{c0}$, where $T_{Gi}$ is defined as the temperature at which the heat capacity given by the mean-field theory is equal to the fluctuation-driven heat capacity~\cite{Larkin}. As is seen, $Gi$ defines the temperature interval near $T_{c0}$, where the pair fluctuations cannot be ignored. For the Q1D+2D system there are two nonzero stiffness coefficients in the GL functional, the effective dimensionality of the GL theory is $2$, and the corresponding $Gi$ number~\cite{Larkin,Salasnich2019} is expressed as
\begin{equation}
Gi^{(2D)}=\frac{T_{c0}b_\psi n_z}{4\pi a^\prime_\psi \sqrt{\mathcal{K}^{(x)}_\psi \mathcal{K}^{(y)}_\psi}},
\label{eq.GN2D_1}
\end{equation}
where $a^\prime_\psi=da_\psi/dT$. Notice that $n_z$ appears in Eq.~(\ref{eq.GN2D_1}) because $N_\nu$ for Q1D and Q2D bands has the dimensions of the 3D DOS, taking into account the degeneracy of the momentum states along the $z$ direction in the Q2D case and along the $y$ and $z$ directions in the Q1D case. The factor $n_y$ is absorbed in the coefficients $b_{\psi},a^\prime_{\psi}$, and $\mathcal{K}^{(\alpha)}_\psi$ whereas $n_z$ is not only included in these coefficients  (through $N_\nu$) but also appears explicitly in Eq.~(\ref{eq.GN2D_1}). As $b_\psi \propto n_z,\, a'_\psi \propto n_z$, and ${\cal K}^{(\alpha)}_\nu \propto n_z$, one can see that $n_z$ does not eventually contribute to $Gi^{2D}$. Then, one can rewrite the right-hand side of Eq.~(\ref{eq.GN2D_1}) in the form of the standard 2D Ginzburg number~\cite{Larkin, Salasnich2019} with the coefficients that depend on the band DOSs accounting only the states associated with the $x$ and $y$ directions. Utilizing the expressions for the coefficients $a_\psi,b_\psi$ and ${\cal K}^{(x,y)}_\psi$ given by Eq.~(\ref{eq.coeff-psi}), we arrive at 
\begin{equation}
Gi^{2D}=Gi_1^{2D}\frac{b_2/b_1+S^4}{S \big(a'_2/a'_1+S^2\big)\sqrt{{\cal K}^{(x)}_2/{\cal K}^{(x)}_1+S^2}},
\label{eq.GN2D_2}
\end{equation}
where the Ginzburg number of band $1$ is given by 
\begin{equation}
Gi_1^{2D}=\frac{T_{c0}b_1 n_z}{4\pi a^\prime_1\sqrt{\mathcal{K}^{(x)}_1 \mathcal{K}^{(y)}_1}}=
\frac{T_{c0}}{\mu+|\varepsilon_0|},
\end{equation}
where $a^\prime_\nu=da_\nu/dT$ and Eqs.~(\ref{eq.coeff1-ab}) and (\ref{eq.coeff1-K2D}) are used. 

Similarly, for the Q1D+3D case there are three nonzero stiffness coefficients in the GL functional of the two-band system, the number of the effective dimensions of the GL functional (\ref{eq.free4}) is $3$, and the Ginzburg number is expressed in the form~\cite{Larkin,Saraiva2020} 
\begin{equation}
Gi^{3D}=\frac{1}{32\pi^2}\frac{T_{c0}b_\psi^2}{a^\prime_\psi
{\cal K}^{(x)}_\psi {\cal K}^{(y)}_\psi {\cal K}^{(z)}_\psi}.
\label{eq.GN3D_1}
\end{equation}
Utilizing Eq.~(\ref{eq.coeff-psi}), one finds
\begin{equation}
Gi^{3D}=Gi_1^{3D}\frac{(b_2/b_1+S^4)^2}{S^4 (a'_2/a'_1+S^2) \big(\mathcal{K}^{(x)}_2/\mathcal{K}^{(x)}_1+S^2\big)},
\label{eq.GN3D_2}
\end{equation}
where the Ginzburg number of band $1$ with the 3D dispersion is given by
\begin{equation}
Gi_1^{3D}=\frac{1}{32\pi^2}\frac{T_{c0}b_1^2}{a^\prime_1
\mathcal{K}^{(x)}_1 \mathcal{K}^{(y)}_1 \mathcal{K}^{(z)}_1},
\label{eq.GN3D_3}
\end{equation}
which can be rewritten as
\begin{equation}
Gi_1^{3D}=\frac{27\pi^4}{14\zeta(3)}\left(\frac{T_{c0}}{\mu+|\varepsilon_0|}\right)^4,
\label{eq.GN3D_4}
\end{equation}
see Eqs.~(\ref{eq.coeff1-ab}) and (\ref{eq.coeff1-K3D}). 

It is instructive to examine the limiting cases $S \to 0$ and $S \to \infty$. As is mentioned above, when $S \to \infty$, band $2$ does not contribute [see Eq.~(\ref{eq.eta})], and superconductivity is determined by the condensate in band $1$. In this limit, Eqs.~(\ref{eq.GN2D_2}) and (\ref{eq.GN3D_2}) yield, respectively, $Gi^{2D} \to Gi_1^{2D}$ and $Gi^{3D} \to Gi_1^{3D}$. 

In the opposite limit $S \to 0$, the contribution of band $1$ is negligible and $Gi \to \infty$ for both Q1D+2D and Q1D+3D systems. Formally, the divergence in the Ginzburg number follows from the fact that ${\cal K}^{(y,z)}_\psi = S^2{\cal K}^{(y,z)}_1 \to 0$ at $S \to 0$ for the Q1D+3D system in Eq. (\ref{eq.GN3D_1}) and ${\cal K}^{(y)}_\psi = S^2{\cal K}^{(y)}_1 \to 0$ for the 1D+2D system in Eq.~(\ref{eq.GN2D_1}). This implies that for one or two spatial directions the integral over the momentum, appearing in the expression for the fluctuation-driven heat capacity [see Ref.~\cite{Larkin}], becomes divergent. This divergence is artificial and has to be regularized, e.g. by introducing the momentum cut-off in the integration (i.e., by taking account of the boundary of the Brillouin zone). However, since we are interested in the regime of the pair-fluctuation suppression with $Gi$ significantly smaller than $1$, we can simply ignore this regularization in our further calculations.

\subsection{Fluctuation corrections to $T_c$}

The Ginzburg number $Gi$ gives a good estimate for the temperature range in the vicinity of the critical temperature where thermal fluctuations are large. However, the fluctuation suppress the superconductivity and reduce the critical temperature itself. As the thermal fluctuations of the two-band system are controlled by the effective single-component GL free-energy functional, we can utilize standard expressions connecting the Ginzburg number with the fluctuation-driven shift of the critical temperature in single-band superconductors, see details in Ref.~\onlinecite{Saraiva2020}. For the case of the GL functional with the two effective dimensions (Q1D+Q2D), one can use 
\begin{equation}
\frac{\delta T_c}{T_c}=4Gi^{(2D)}, \quad \delta T_c=T_{c0}-T_c,
\label{eq.shift2D}
\end{equation}
where $T_c$ is the Berezinski-Kosterlitz-Thouless (BKT) transition temperature~\cite{Kosterlitz1973} [see the discussion in Ref.~\onlinecite{Salasnich2019}]. One can also apply the renormalization group result~\cite{Larkin, Salasnich2019}, which yields $\delta T_c/T_c=2Gi^{(2D)}\ln(1/4Gi^{(2D)})$. However, this formula is applicable only when~\cite{Salasnich2019} $\delta T_c/T_c \lesssim 0.1$~(one can see that its right-hand side even changes sign for $Gi > 1/4$). In addition, values of the critical temperature calculated within the BKT scenario of the pair fluctuations are very close to the renormalization group estimates for $\delta T_c/T_c=0.005\div 0.1$. The difference becomes significant only when $\delta T_c/T_c \lesssim 0.001$ but in this regime $T_c$ and $T_{c0}$ are almost indistinguishable. Thus, it is more convenient to choose the BKT variant (\ref{eq.shift2D}) in our present study. 

For the Q1D+3D system we have three nonzero stiffness coefficients and can utilize 
\begin{equation}
\frac{\delta T_c}{T_c}=\frac{8}{\pi}\sqrt{Gi^{(3D)}}, 
\label{eq.shift3D}
\end{equation}
which is the result of the 3D renormalization group analysis~\cite{Larkin}.

Thus, employing the Ginzburg number calculated with Eqs.~(\ref{eq.shift2D}) and (\ref{eq.shift3D}) for the Q1D+Q2D and Q1D+3D models, respectively, one can find the related shift of the critical temperature (with respect to the mean-field superconductive critical temperature) from Eqs.~(\ref{eq.shift2D}) and (\ref{eq.shift3D}).

\section{Numerical results}
\label{sec3}

Now we investigate how the critical temperature $T_c$ depends on the system parameters, including the dimensionality and the depth of the higher-dimensional band, as well as on the interaction strength and effective carrier masses of the both contributing bands. The analysis is done in two steps: first, we calculate the mean-field critical temperature $T_{c0}$ and then the critical temperature $T_{c}$, renormalized due to the pair thermal fluctuations.  

\subsection{Model parameters}

The mean-field critical temperature $T_{c0}$ is determined by the three dimensionless coupling constants $\lambda_{11}$, $\lambda_{22}$, $\lambda_{12}$ and the chemical potential $\mu$. Within the adopted model, $\mu \sim 0$ determines the proximity to the Lifshitz transition point (the Feshbach-like resonance) and is used as a variable in our calculations. We avoid the trivial regime of a dominant higher-dimensional band $\lambda_{11} \gg \lambda_{22}$ at which the system characteristics are close to those of a conventional BCS superconductor. Notice that similar situation occurs for large pair-exchange couplings $\lambda_{12} \gg \lambda_{22}$ because intensive Cooper-pair transfer between the contributing bands washes out the Q1D effects. Thus, the most physically interesting is the case of $\lambda_{11},\lambda_{12} \lesssim \lambda_{22}$, where the Q1D physics is still important. 

Below we choose $\lambda_{22}=0.2$, which is in the range of the values typical for the dimensionless couplings of conventional superconductors~\cite{Fetter}. For the higher-dimensional band, we investigate two variants: the vanishing coupling constant $\lambda_{11} = 0$ ($\lambda_{11} \ll \lambda_{22}$), and $\lambda_{11} = 0.24$ ($\lambda_{11} \simeq \lambda_{22}$). Finally, to study the effects of the interband interactions on the system properties, several values of the pair-exchange coupling $\lambda_{12}$ are considered.  

To calculate the critical temperature renormalized by the thermal pair fluctuations, one needs to know the Ginzburg number of the two-band system in addition to $T_{c0}$. The former depends on the Ginzburg number of the higher-dimensional band $Gi_1$, the band occupation parameter $S$, as well as on the ratios $N_2/N_1$ and $v_2/v_1$. As follows from Eq.~(\ref{eq.S}), $S$ is controlled by $\lambda_{\nu\nu'}$, $\mu$, and $N_2/N_1$, whereas $Gi_1$ is determined by the chemical potential $\mu$ and the band depth $\varepsilon_0$. Further, the ratio $v_2/v_1= \sqrt{m_1/ \big(m_2(\tilde{\mu} + |\tilde{\varepsilon}_0|) \big)}$ depends on $\mu$, $\varepsilon_0$, and the band mass ratio $m_2/m_1$. Consequently, we find that  $T_c$ is governed by $\lambda_{\nu\nu'}$, $\mu$, $\varepsilon_0$, $N_2/N_1$, and $m_2/m_1$. Notice that there is no need to specify the cut-off energy $\hbar\omega_c$ as the latter simply determines the energy scale of the model.  

\begin{figure}[t]
\centering
\includegraphics[width=1.0\linewidth]{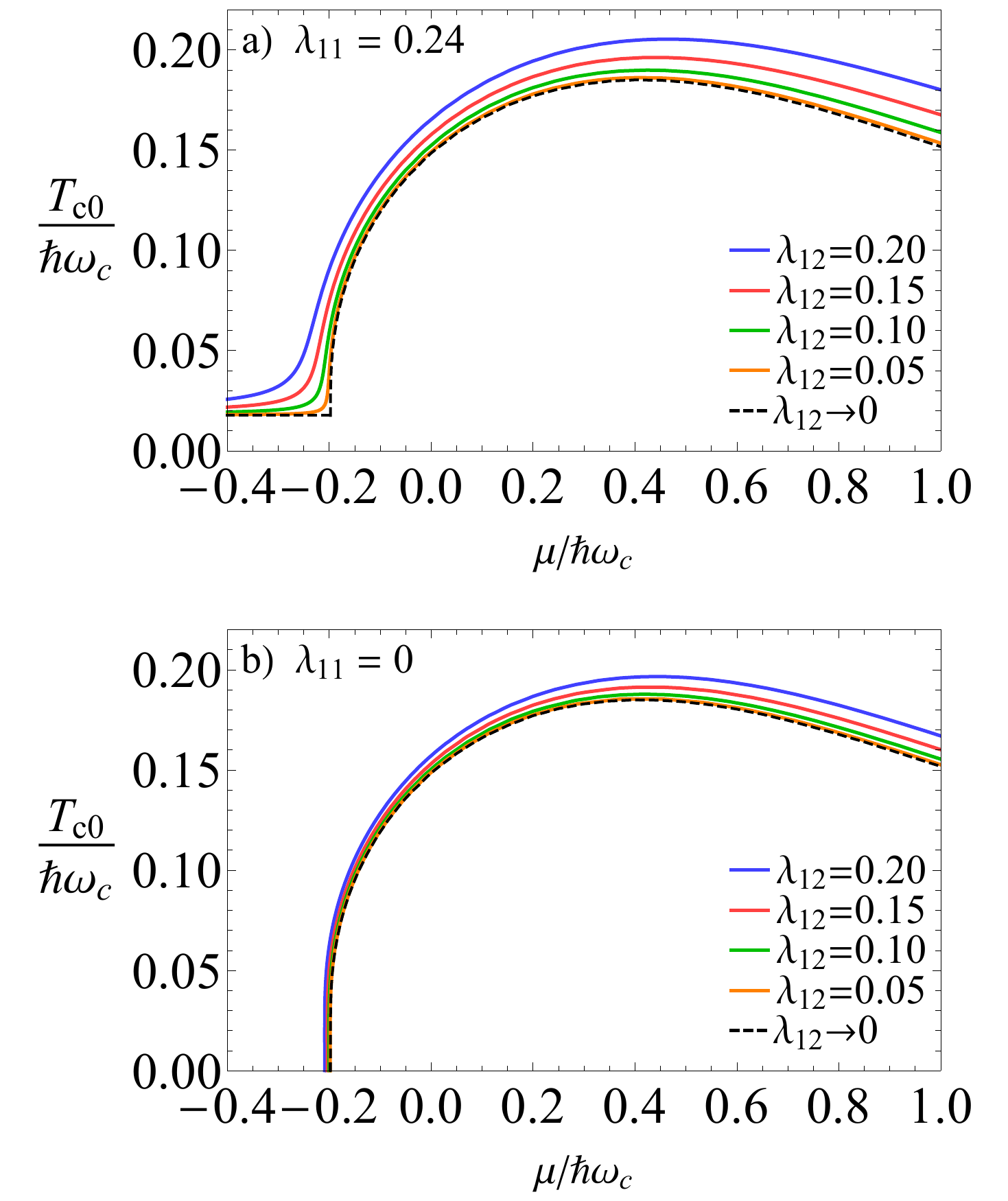}\\
\caption{\label{fig2} The mean-field critical temperature $T_{c0}$ of the two-band system as a function of the chemical potential $\mu$. Panels a) and b) demonstrate results obtained for the intraband couplings $\lambda_{11}=0.24$ and $\lambda_{11}=0$ in the higher-dimensional band, results for the Q1D+Q2D and Q1D+3D models are the same.  The calculations assume the intraband coupling in the Q1D band is $\lambda_{22}=0.2$ while the pair-exchange interband coupling is $\lambda_{12} = 0,0.05,0.1,0.15,0.2$. The dotted line represents $T_{c0}$ obtained in the limit $\lambda_{12} \to 0$ of uncoupled bands.}
\end{figure}

The ratio of the band DOSs $N_2/N_1$ is close to $1$ in most of two-band superconductors, see Ref.~\onlinecite{Salasnich2019}, and for simplicity we choose $N_2/N_1=1$. Finally, we consider different values of $\varepsilon_0$ and $m_2/m_1$, to investigate how the depth of band $1$ impacts $T_c$ and how this impact is sensitive to the band mass ratio $m_2/m_1$. In multiband superconductors the effective band masses can significantly deviate from that of free electrons~\cite{Kasahara2014}, and therefore the ratio $m_2/m_1$ is not necessarily equal to $1$. 

\subsection{Mean-field critical temperature $T_{c0}$}

Figure~\ref{fig2} demonstrates $T_{c0}$ versus the chemical potential $\mu$ for $\lambda_{11}=0.24$ in panel a) and for $\lambda_{11}=0$ in panel b); the results of solving Eq.~(\ref{eq.Tc0}) are given for the set of the pair-exchange couplings $\lambda_{12}=0,0.05, 0.1, 0.15, 0.2$. Notice that the mean-field critical temperature does not depend on the number of the dimensions of band $1$. Moreover, one can see that $T_{c0}$ is not much sensitive to particular values of $\lambda_{11}$ and $\lambda_{12}$ at $\mu > -0.2$~(here and below the values of the energy related quantities are given in units of the cut-off energy). The mean-field critical temperature enhancement is mainly determined by the van Hove singularity of the Q1D DOS, which results in the Q1D Feshbach-like resonance enhancement of superconductivity and the related enhancement of $T_{c0}$. Notice that this enlargement starts at $\mu \simeq -0.2$, not at $\mu =0$. This downward shift is related to the Cooper-pair binding energy $\sim {\rm max}[T_{c0}] \simeq 0.2$ and to the temperature-dependent smearing of the Fermi surface of the normal state. One can see that only at $\mu < -0.2$ the contribution of the Q1D band is negligible, and $T_{c0}$ is fully determined by the higher-dimensional band: $T_{c0}=0$ for $\lambda_{11}=0$ and $T_{c0} \approx 0.02$ for $\lambda_{11}=0.24$. We note that even for $\lambda_{11}=0.24$ the maximal value of $T_{c0}$ is larger by an order of magnitude than $T_{c0}$ at $\mu < -0.2$. We note that the possibility of a significant increase of $T_{c0}$ is at the core of the researchers interest to the Feshbach-like resonances in multiband superconductors~\cite{Innocenti2010, Bianconi2013,Mazziotti2017}. 

\subsection{Renormalized $T_c$ close to the Lifshitz transition}

\begin{figure}[t]
\centering
\includegraphics[width=1.0\linewidth]{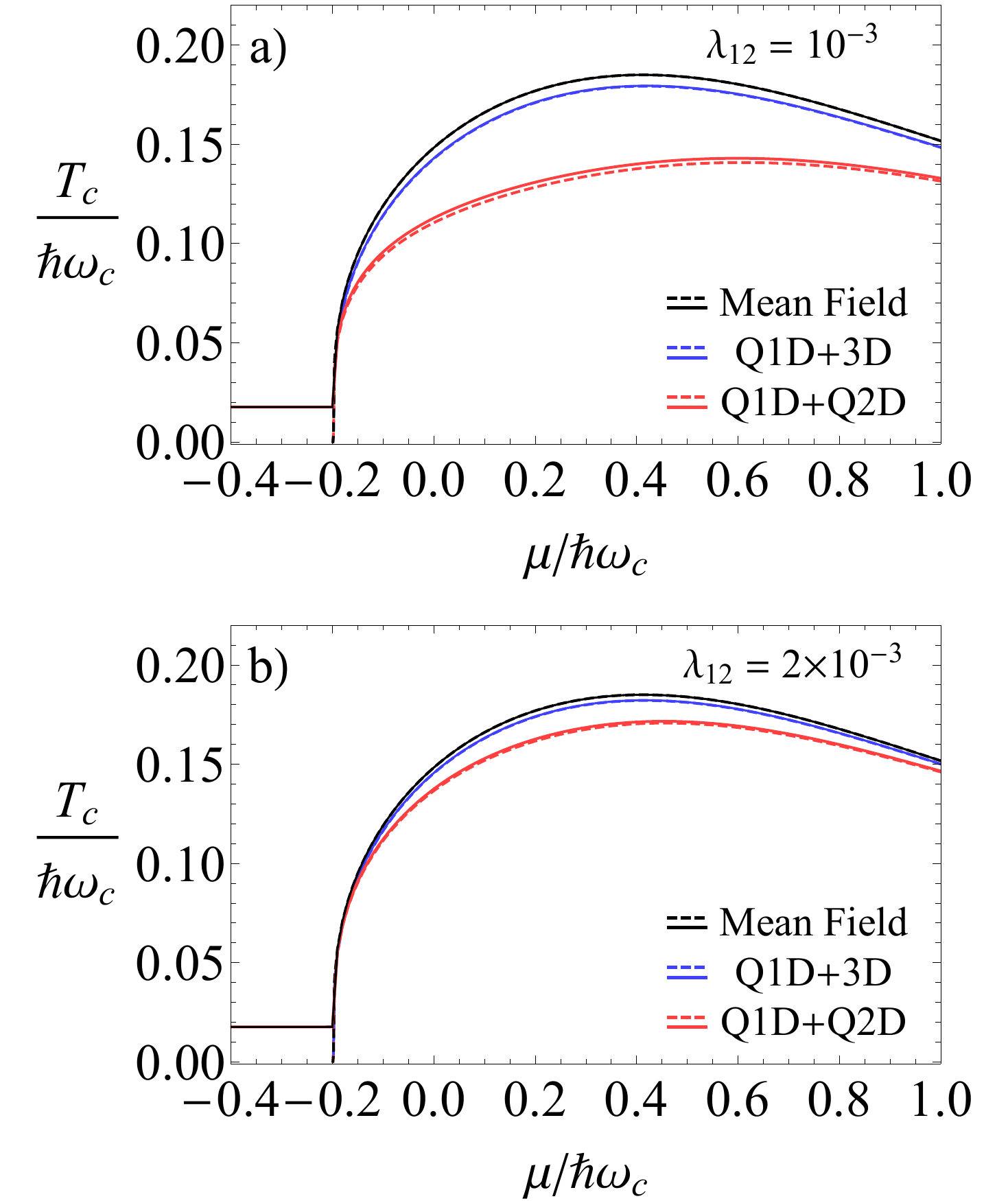}
\caption{\label{fig3}
The critical temperature $T_c$ renormalized by the pair fluctuations versus $\mu$, calculated for $\lambda_{12}=2\cdot 10^{-3}$ (panel a) and $\lambda_{12}= 5\cdot 10^{-3}$~(panel b). The depth of the higher-dimensional band is chosen as $|\varepsilon_0| =300\hbar\omega_c$ and the mass ratio is set to $m_2/m_1=1$. Results for the Q1D+Q2D and Q1D+3D models are given by blue and red lines, respectively. The dotted lines correspond to $\lambda_{11}=0$ whereas the solid lines represent the case of $\lambda_{11}=0.24$. Black lines give the mean-field critical temperature, for comparison.}
\end{figure}

We now consider impact of the thermal pair fluctuations and calculate the critical temperature $T_c$ renormalized by the fluctuations. Figure \ref{fig3} shows $T_c$ as a function of $\mu$, calculated for the pair-exchange couplings $\lambda_{12} = 10^{-3}$~(panel a) and $\lambda_{12} = 2\times 10^{-3}$~(panel b). Here the depth of band $1$ with respect to the bottom of band $2$ is chosen as $|\varepsilon_0|= 300$, which yields the Fermi energy of the higher-dimensional band $E_F= |\varepsilon_0| + \mu \approx 300$. This value of the Fermi energy is close to that of the conventional elemental superconductors, see Ref.~\onlinecite{deGennesBook}~(e.g., for ${\rm Al}$ we have $E_F \approx 350$ whereas in ${\rm Pb}$ one gets $E_F \approx 1000$). The ratio of the band masses is chosen as $m_2/m_1=1$. Results for the Q1D+Q2D and Q1D+3D systems are shown by the red and blue lines, correspondingly. To demonstrate the shift of the critical temperature by the fluctuations, $T_{c0}$ is also shown Figs.~\ref{fig3} a) and b) by the black lines. In all cases, solid lines correspond to $\lambda_{11}=0.24$ and the dotted lines are related to $\lambda_{11}=0$. 

Figure~\ref{fig3} demonstrates that the pair fluctuations are negligible for $\mu < 0.2$ and important for $\mu > 0.2$, where the contribution of the Q1D band matters. It is seen from Fig.~\ref{fig3} a) that for $\mu > 0.2$ the difference between the results for $\lambda_{11} =0.24$ and $\lambda_{11}=0$ is almost negligible for the Q1D+Q2D system and even not visible for the Q1D+3D case. Moreover, as it follows from Fig.~\ref{fig3} b), this difference tends to disappear for larger values of $\lambda_{12}$. For example, it is not visible in panel b) for both the Q1D+Q2D and Q1D+3D systems. Thus, though the higher-dimensional condensate can exist only due to the Cooper-pair transfer from the Q1D band at $\lambda_{11} =0$, the coupling to the higher-dimensional band ``kills" the Q1D pair fluctuations similarly to the case of a finite value of $\lambda_{11}$. 

As is well known, the role of fluctuations increases in low dimensional samples~\cite{Larkin}. Then, one can expect that the effect of the pair fluctuations on the critical temperature in the Q1D+Q2D model should be significantly stronger than that in the Q1D+3D model. Figure~\ref{fig3} demonstrates that $T_c$ is indeed lower in the Q1D+Q2D system. However, the most pronounced difference between the critical temperatures of the two models is only about $30\%$, see the results in Fig.~\ref{fig3} a) for $\mu \approx 0.2$-$0.6$. The reason for such a weak dependence of $T_c$ on the dimensionality of band $1$ is originated in the dependence of $\delta T_c/T_c$ on $Gi$: the fluctuation-driven shift of the critical temperature is linear in $Gi^{2D}$ while it is proportional to the square root of $Gi^{3D}$~[cf. Eqs.~(\ref{eq.shift2D}) and (\ref{eq.shift3D})]. Though $Gi^{2D}$ is indeed by orders of magnitude larger than $Gi^{3D}$, the difference between the corresponding values of $\delta T_c$ is much less pronounced. For instance, at $\mu \approx 0.4$ and $\lambda_{12}=10^{-3}$ we have $Gi^{2D} \approx 5 \times 10^{-2}$ and $Gi^{3D} \approx 4.5\times 10^{-4}$.

The most important result is that $T_c$ rapidly approaches $T_{c0}$ when the interband coupling increases above $\lambda_{12} \gtrsim  10^{-3}$ [Fig. \ref{fig3}]. Notice, that in this regime one can still have $\lambda_{12} \ll \lambda_{22}$, so that at $\mu > - 0.2$, $T_{c0}$ is determined by the resonant Q1D band. This conclusion is independent on the intraband coupling $\lambda_{11}$ in the higher-dimensional band and holds even when the higher-dimensional condensate appears only due to the proximity-like effect between the bands.  

\subsection{Impact of the energy depth of the reservoir band}

\begin{figure}[t]
\centering
\includegraphics[width=0.95\linewidth]{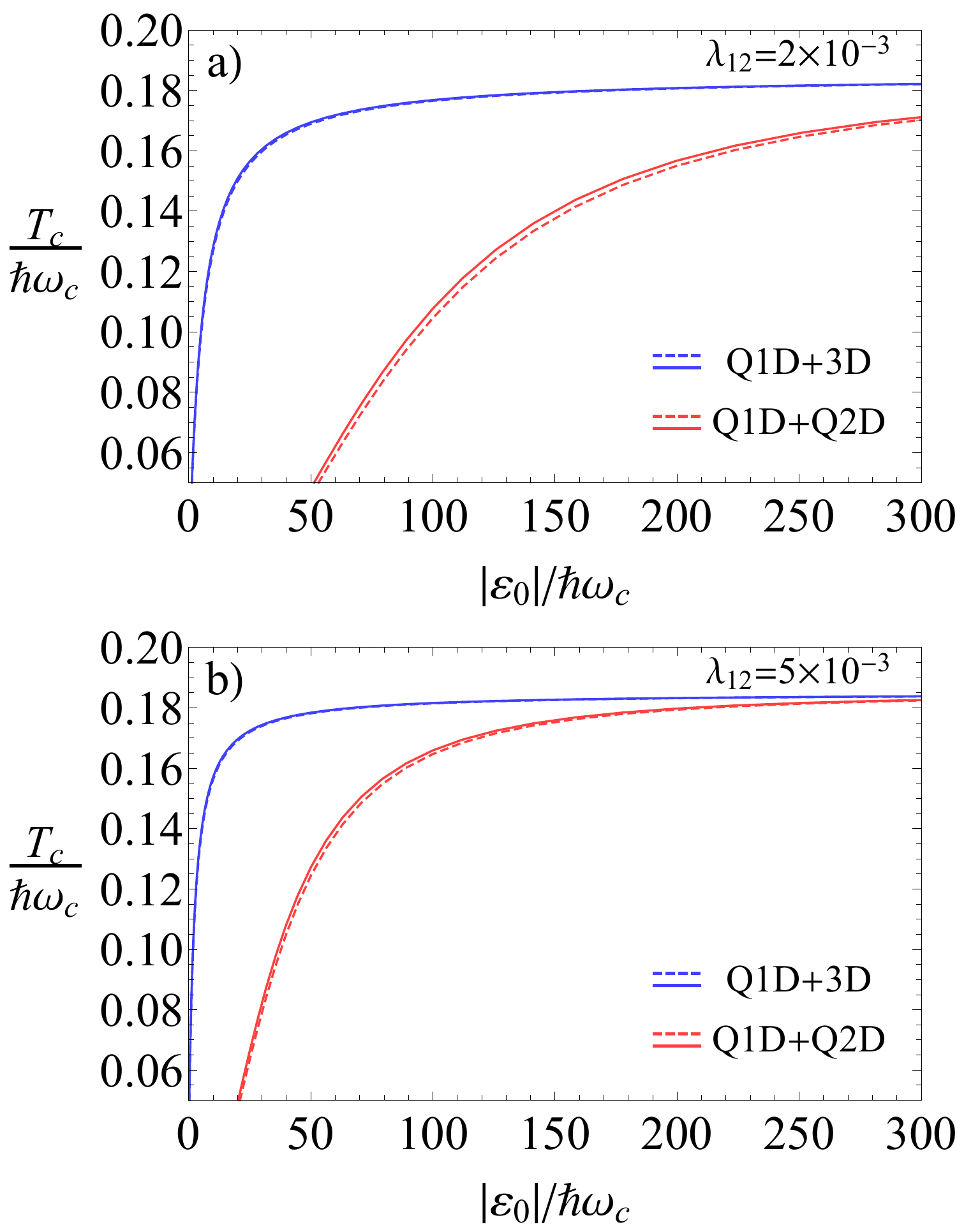}\\
\caption{\label{fig4}
The critical temperature $T_c$ renormalized by the fluctuations, plotted as a function of the reservoir band depth $|\varepsilon_0|$, calculated for $\mu/\hbar\omega_c=0.4$ and $m_2/m_1=1$: panel a) represents $\lambda_{12}=2\cdot 10^{-3}$ and panel b) ~corresponds to $\lambda_{12}= 5\cdot 10^{-3}$. The results for the Q1D+Q2D (blue lines) and Q1D+3D (red lines) systems are shown for both the passive ($\lambda_{11}=0$) and active ($\lambda_{11}=0.24$) regimes of band $1$ by the dotted and solid lines.}
\end{figure}

The results in Fig.~\ref{fig3} are obtained for $|\varepsilon_0|=300$~(recall that all energy-related quantities are given in units of the cut-off energy in the text) and, as is mentioned above, the corresponding Fermi energy of band $1$ is in the range of the Fermi energies of the conventional elemental superconductors. However, in novel superconducting materials $E_F$ can be significantly smaller, down to $10$ or even below this value~\cite{Lubashevsky2012, Okazaki2013, Coldea2018, Hanaguri2019}. This is why in Fig.~\ref{fig4} we consider how $T_c$ is sensitive to $|\varepsilon_0|$. We again employ $\lambda_{22}=0.2$, $\lambda_{11}=0$ and $0.24$~(dotted and solid lines, respectively) and use $\lambda_{12}=2\cdot 10^{-3}$ in Fig.~\ref{fig4} a) and $\lambda_{12}=5 \cdot 10^{-3}$ in Fig.~\ref{fig4} b). The chemical potential is now fixed as $\mu=0.4$ which corresponds to a nearly maximal value of $T_c$, see  Fig.~\ref{fig3} b). Here we still adopt $m_2/m_1=1$ while this value will be changed below.  

Our results demonstrate that $T_c$ increases with $|\varepsilon_0|$, asymptotically approaching $T_{c0}$ (saturation) for very large values of $|\varepsilon_0|$. Notice, that $T_{c0}$ is only weakly dependent on $\lambda_{11}$ and $\lambda_{12}$, as shown in Figs.~\ref{fig2} and \ref{fig3}. Thus, we find that the pair fluctuations are quenched in the limit $|\varepsilon_0| \to \infty$, which is in agreement with our previous results for the two-band model with a deep higher-dimensional band~\cite{Saraiva2020}.  

For relatively small values of $|\varepsilon_0|$ one finds that $T_c$ deviates notably from $T_{c0}$. In this case the coherence length of band $1$ decreases, which leads to an increase of $Gi_1$ and so, to an enhancement of the pair fluctuations in the reservoir band, see Ref.~\onlinecite{Larkin}. This, in turn, results in a rise of $Gi$ and $\delta{T_c}$, see Eqs.~(\ref{eq.GN2D_2}), (\ref{eq.GN3D_2}), (\ref{eq.shift2D}), and (\ref{eq.shift3D}). In addition, the ratio $v_2/v_1$ decreases with decreasing $|\varepsilon_0|$ and so does the ratio ${\cal K}^{(x)}_2/{\cal K}^{(x)}_1$. As the latter appears in the denominators of Eqs.~(\ref{eq.shift2D}) and (\ref{eq.shift3D}), one obtains an additional contribution to an increase of $Gi$ and $\delta{T_c}$. 

One can see that the suppression of the fluctuations is more effective in the Q1D+3D system but again, we do not observe an order of magnitude difference between the critical temperatures in the Q1D+Q2D and Q1D+3D models even at small $|\varepsilon_0|$. Notice that only the results for $T_c > 0.05$ are shown in Fig.~\ref{fig3} since our formalism does not apply for strong pair fluctuations with $Gi$ close to $1$, see the discussion after Eq.~(\ref{eq.GN3D_4}). 

\begin{figure}[t]
\centering
\includegraphics[width=0.95\linewidth]{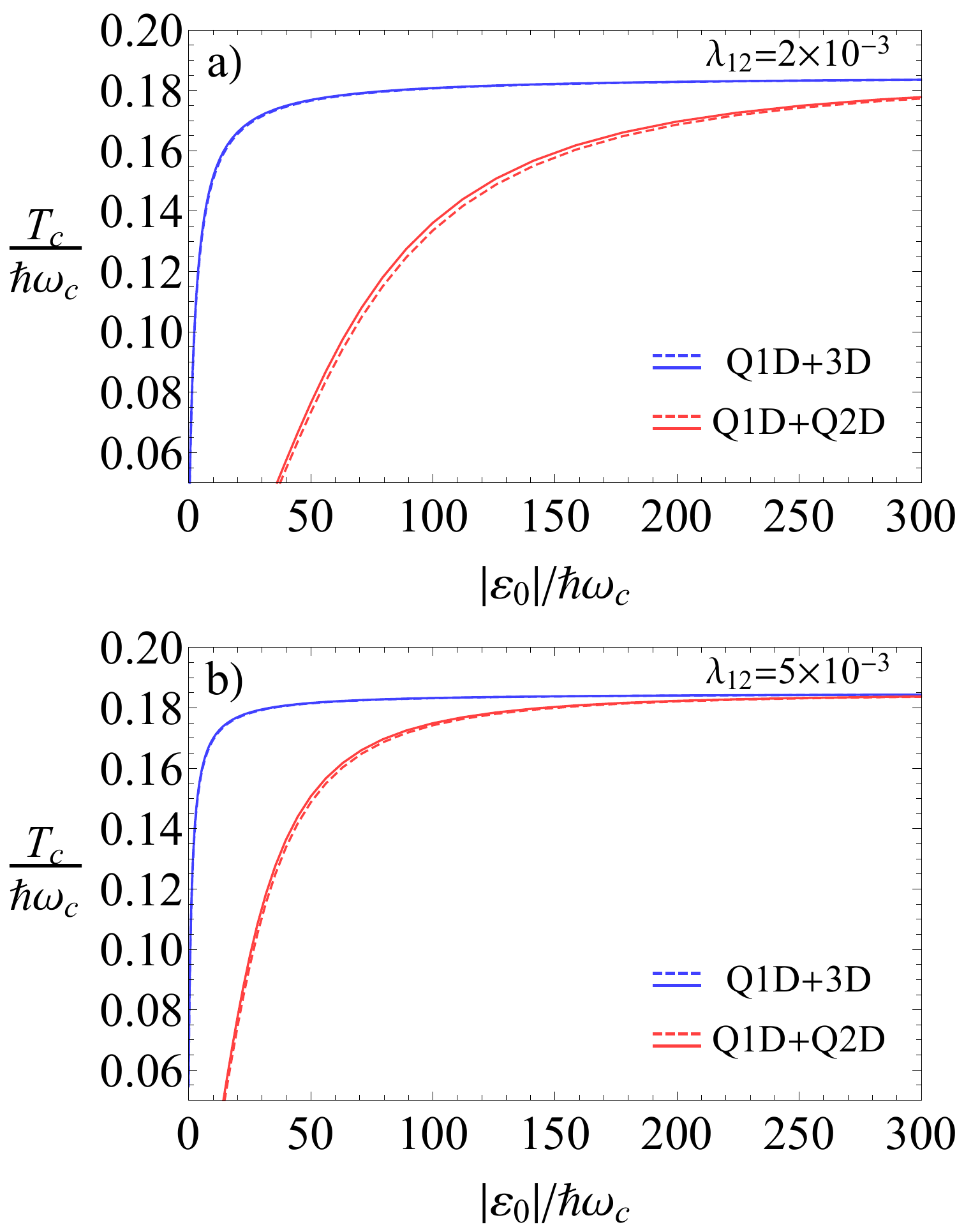}\\
\caption{\label{fig5}
The same as in Fig.~\ref{fig4} but for $m_2/m_1=1/4$.}
\end{figure}

The effective band masses can significantly deviate from the free electron mass in multiband superconductors~\cite{Kasahara2014} and then the ratio $m_2/m_1$ can be notably different from $1$. As is mentioned above, the value of $v_2/v_1$ is controlled by $|\varepsilon_0|$ and by the ratio $m_2/m_1$. The larger (smaller) is the value of $m_2/m_1$, the smaller (larger) is the ratio $v_2/v_1$ at a given $|\tilde{\varepsilon}_0|$ and, then, the larger (smaller) is the impact of the pair fluctuations. For illustration, we calculate $T_c$ with the same microscopic parameters as in Fig.~\ref{fig4} but for $m_1=4m_2$, see Fig.~\ref{fig5}. The results in Fig.~\ref{fig5} are similar to those in Fig.~\ref{fig4} but the region of strong fluctuations shifts to lower values of $|\varepsilon_0|$. For instance, for the Q1D+3D system $T_c$ is equal to a half of $T_{c0}$ at $|\tilde{\varepsilon}_0| \approx 7$-$8$ in panel a) and at $|\tilde{\varepsilon}_0| \approx 2$-$3$ in panel b).

Thus, a drop of $T_c$ at small $|\varepsilon_0|$ demonstrates that the Q1D thermal pair fluctuations are significantly enhanced when the reservoir band approaches its shallow regime. However, quite surprisingly, we find that the fluctuations are significantly weakened even when the Fermi energy of the higher-dimensional band is by two orders of magnitude smaller than $E_F$ in the conventional elemental superconductors. 

\section{Conclusion}
\label{sec4}

Although the mean-field critical temperature of a single-band Q1D superconductor can be very large when approaching the van Hove singularity, the thermal pair fluctuations suppress or even eliminate altogether the superconductivity, reducing the critical temperature to very low values or to zero. However, the situation changes dramatically when a Q1D band is coupled to the reservoir condensate of a higher-dimensional band due to the pair transfer between the bands. This pair-exchange coupling can effectively suppress the detrimental Q1D pair thermal fluctuations~\cite{Saraiva2020}. 

This work studies details of how thermal superconducting fluctuations are quenched in a two-band system comprising a Q1D and Q2D/3D bands, and the critical temperature $T_c$ approaches its mean-field value $T_{c0}$. The focus of the study is to clarify how the screening of the pair thermal fluctuations in such a two-band system depends on the microscopic parameters specifying the single-particle bands and the intra- and interband pairing interactions. 

Although the two-band system is controlled by a fairly large number of different parameters, our calculations demonstrate that the fluctuation suppression effect is a general phenomenon, taking place in a wide parametric domain. This domain for the Q1D+Q2D model is very close to that of the Q1D+3D model. Furthermore, we find that the fluctuation suppression occurs even if the reservoir band alone does not develop the superconducting state and also when the reservoir band is nearly shallow, having a relatively small band Fermi energy in comparison to that of the conventional metallic superconductors. Notice that the case of unusually small Fermi energies in superconducting compounds is not a theoretical assumption or oversimplification, it is relevant, e.g., for FeSe, where multiple overlapping bands crossing the Fermi level have similar small depths~\cite{Coldea2018,Hanaguri2019}. 

The present results have both academic and practical importance. Our work uncovers an important aspect of the physics of multiband superconductors exhibiting a BCS-BEC crossover (see experimental results in Refs.~\onlinecite{Kasahara2014, Lubashevsky2012, Okazaki2013, Coldea2018, Hanaguri2019} and theoretical studies in Refs.~\onlinecite{Perali1996, Bianconi1997, Shanenko2006, Shanenko2010, Innocenti2010, Chen2012, Bianconi2013, Mazziotti2017}). It is commonly expected that thermal superconducting fluctuations proliferate when a system approaches the BCS-BEC crossover and then goes into the BEC regime. In contrast, our study and investigations of the previous works~\cite{Salasnich2019,Saraiva2020} demonstrate that those fluctuations, being detrimental in the single-band case, are screened in the multiband superconductors by the pair-exchange coupling of a shallow-band condensate in the BCS-BEC-crossover regime to the band that is still in the BCS regime. As a result, the preformed Cooper pairs induced by thermal fluctuations disappear and the critical temperature of the global coherence approaches the pair-formation temperature. Interestingly, this conclusion agrees with the recent scanning tunnelling microscopy results for FeSe~\cite{Hanaguri2019}.

From the practical point of view, our results are encouraging to merit engineering and further detailed investigations of the Q1D multiband superconductors, such as recent materials A$_2$Cr$_3$As$_3$, with A = K, Rb, Cs~\cite{Jiang2015, Bao2015, Zhi2015, Tang2015A, Tang2015B, Wu2019, Xu2020}. Our investigation confirms that tuning the Lifshitz topological transition associated with the edge of a Q1D band, e.g. by means of doping, applying external pressure or chemical engineering, is a very promising way to achieve robust high-$T_c$ superconductivity.

\begin{acknowledgements}
The study has been funded within the framework of the HSE University Basic Research Program.
\end{acknowledgements}

\appendix
\section{Coefficients for the Q1D GL theory}
\label{secApp}

In this Appendix we derive the coefficients for the Q1D GL theory for the reader convenience. Below we follow the standard procedure of the microscopic derivation of the GL formalism introduced by Gor'kov~\cite{Gorkov1959, AGD1965}.  In vicinity of the mean-field critical temperature the gap function $\Delta_{\nu}({\bf r})$ is small and the corresponding anomalous Green function $R_{\nu}({\bf r})$ can be represented by series in powers of $\Delta_{\nu}({\bf r})$.  Adopting the Gor'kov truncation procedure~\cite{Gorkov1959,AGD1965} and keeping only the lowest nonlinear term, we obtain the anomalous Green function as
\begin{align}
R_{\nu}({\bf r})={\cal I}_{a\nu}[\Delta_{\nu}({\bf r})] +{\cal I}_{ b\nu}[\Delta_{\nu}({\bf r})],
\label{eq.gapexp}
\end{align}
where
\begin{align}
{\cal I}_{a\nu}=\int d^3{\bf r}' K_{a\nu}({\bf r},{\bf r}')\Delta_{\nu}({\bf r}')
\label{eq.Ia2}
\end{align}
and 
\begin{align}
{\cal I}_{b\nu}=&\int d^3{\bf r}'d^3{\bf r}''d^3{\bf r}''' K_{b\nu}({\bf r},{\bf r}',{\bf r}'',{\bf r}''')\nonumber\\
&\times \Delta_{\nu}({\bf r}')\Delta_{\nu}^\ast({\bf r}'')\Delta_{\nu}({\bf r}'''),
\label{eq.Ib2}
\end{align}
with the integral kernels defined by
\begin{align}
K_{a\nu}({\bf r},{\bf r}') = -T\sum_\omega\mathcal{G}_{\nu \omega}^{(0)}({\bf r},{\bf r}')\bar{\mathcal{G}}_{\nu\omega}^{(0)}({\bf r},{\bf r}')
\end{align}
and
\begin{align}
K_{b\nu}({\bf r},{\bf r}',{\bf r}'',{\bf r}''')=&-T\sum_\omega \mathcal{G}_{\nu\omega}^{(0)}({\bf r},{\bf r}')
\bar{\mathcal{G}}_{\nu\omega}^{(0)}({\bf r}',{\bf r}'')\nonumber\\
&\times \mathcal{G}_{\nu\omega}^{(0)}({\bf r}'',{\bf r}''')\bar{\mathcal{G}}_{\nu\omega}^{(0)}({\bf r}''',{\bf r}). 
\end{align}
Here the normal-state Green function is expressed in terms of the single electron energy $\xi_{{\bf k}\nu}$ as
\begin{equation}
\mathcal{G}_{\nu\omega}^{(0)}({\bf r},{\bf r}')=\int \frac{d^3{\bf k}}{(2\pi)^3} \frac{e^{-i{\bf k}({\bf r}-{\bf r}')}}{i\hbar\omega-\xi_{\bf k\nu}},
\end{equation}
and $\bar{\mathcal{G}}_{\nu\omega}^{(0)}({\bf r},{\bf r}')=-\mathcal{G}^{(0)}_{\nu,-\omega}({\bf r}',{\bf r})$. The integral kernels involve, as usual, the summation over the fermionic Matsubara frequencies $\omega \to \omega_n=\pi T(2n+1)/\hbar$~($n=0,\pm1,\pm2,\ldots$).

Using Eqs.~(\ref{eq.selfcons}) and (\ref{eq.gapexp}), one obtains the integral equations for the gap functions $\Delta_1({\bf r})$ and $\Delta_2(\bf r)$. In the next step, the obtained integral equations are approximated by partial differential equations via using the gradient expansion $({\bf r}'={\bf r} + {\bf z})$
\begin{align}
\Delta_{\nu}({\bf r}') = \sum\limits_{m=0,1,2,\ldots}\frac{({\bf z} \cdot \boldsymbol{\nabla})^m}{m!} \Delta_{\nu}({\bf r}),
\label{eq.grad}
\end{align}
where we keep only the contributions up to second-order spatial derivatives.  Substituting this expansion into the integral gap equations, we obtain  
\begin{align}
{\cal I}_{a\nu}={\cal I}_{a\nu}' + {\cal I}_{a\nu}'',
\label{eq.first}
\end{align}
with
\begin{align}
{\cal I}_{a\nu}'=\int d^3{\bf r}'K_{a2}({\bf r},{\bf r}') \Delta_2({\bf r})
\label{eq.first'}
\end{align}
and
\begin{align}
{\cal I}_{a\nu}''=\int d^3{\bf r}'K_{a2}({\bf r},{\bf r}') \frac{({\bf z}\cdot\boldsymbol{\nabla})^2}{2}\;\Delta_2({\bf r}),
\label{eq.first''}
\end{align}
where the contribution of the first-order derivatives vanishes due to the symmetry $K_{a2}({\bf r},{\bf r}')=K_{a2}({\bf r}',{\bf r})$. Then, Eq.~(\ref{eq.first'}) is rewritten in the form
\begin{align}
{\cal I}_{a\nu}' = T \Delta_2
\int\frac{d^3{\bf k}}{(2\pi)^3}\sum_{\omega} \frac{1}{\hbar^2\omega^2+\xi^2_{{\bf k}\nu}},
\end{align}
which is further reduced to 
\begin{align}
{\cal I}_{a\nu}'=\Delta_2\int\frac{d^3{\bf k}}{(2\pi)^3} \frac{\tanh(\xi_{{\bf k}\nu}/2T)}{2\xi_{{\bf k}\nu}}
\label{eq.first'A}
\end{align}
where we use the well-known relation for the summation over the Matsubara frequencies  
\begin{equation}
\sum_\omega\frac{1}{\hbar^2\omega^2+\xi^2_{{\bf k}\nu}}=\frac{\tanh(\xi_{{\bf k}\nu}/2T)}{2T \xi_{{\bf k}\nu}}.
\label{eq.sum1}
\end{equation}

To proceed further, we recall that our analysis focuses on the GL coefficients for the Q1D band. We take into account that the single-particle energy in band $2$ depends only on $k_x$~(the Q1D band). Then, performing the integration over the momentum ${\bf k}$ for band $2$, we find 
\begin{align}
\int\frac{d^3{\bf k}}{(2\pi)^3}=n_y n_z \int \frac{dk_x}{2\pi},
\label{eq.Q1D}
\end{align}
where constants $n_y$ and $n_z$ are introduced to take into account the Brillouin zone boundaries in the $k$ space. Changing the integration variable to the single-particle energy in Eq.~(\ref{eq.first'A}), we arrive at ($\nu=2$)
\begin{align}
{\cal I}_{a\nu}'
=\Delta_2 N_2 \int\limits_{-\tilde\mu}^{+\infty}d\varepsilon\frac{\tanh(\varepsilon/2\tilde{T})}{\varepsilon\sqrt{\varepsilon+\tilde\mu}}\, \theta(1-|\epsilon|),
\end{align}
where $N_2$ is the Q1D DOS at the cut-off energy, see Eq.~(\ref{eq.dos}), $\tilde{T}$ is the temperature in units of the cut-off energy, and the Heaviside step function $\theta(x)$ is introduced to restrict the integration over single-particle states as $|\xi_{{\bf k}2}| < \hbar\omega_c$. The derived expression can be represented as a series of $\tau=1-T/T_{c0}$~($\tau$ is small near $T_{c0}$). Keeping only the leading and next-to-leading terms in this expansion, one arrives at 
\begin{align}
{\cal I}_2'=({\cal A}_2 -a_2)\Delta_2,
\label{eq.first'B}
\end{align}
with 
\begin{align}
{\cal A}_2=  N_2 \int\limits_{-\tilde\mu}^{+\infty}d\varepsilon\frac{\tanh(\varepsilon/2\tilde{T}_{c0})}{\varepsilon\sqrt{\varepsilon+\tilde\mu}} \,\theta(1-|\epsilon|),
\label{eq.A2}
\end{align}
and 
\begin{align}
a_2= -\tau\frac{N_2}{2\tilde T_{c0}} \int\limits_{-\tilde{\mu}}^{+\infty}d\varepsilon \, \frac{\text{sech}^2\big( \varepsilon / 2\tilde T_{c0} \big) }{\sqrt{\varepsilon+\tilde{\mu}}}\, \theta(1-|\epsilon|), 
\label{eq.a2}
\end{align}
Now, when $\tilde{\mu} \leq 1$, Eqs.~(\ref{eq.A2}) and (\ref{eq.a2}) give the expressions for ${\cal A}_2$ and $a_2$ used in the article, see Eqs.~(\ref{eq.dos}) and (\ref{eq.coeff2}).

We turn now to the calculation of ${\cal I}_{a2}''$. It can be rewritten as
\begin{align}
{\cal I}_{a2}''=\sum\limits_{i,j=1,2,3} \partial_i\partial_j
\Delta_2\int d^3{\bf r}'K_{a2}({\bf r},{\bf r}')\frac{z_i z_j}{2}\, ,
\end{align}
where $z_i$ is the Cartesian component of ${\bf z}$~($i=x,y,z$). The integral in the right-hand side of this expression is rearranged to get 
\begin{align}
{\cal I}_{a2}''= & -\frac{T}{2}\sum\limits_{i,j=1,2,3}\partial_i\partial_j\Delta_2\sum_\omega\int \frac{d^3{\bf k}}{(2\pi)^3}\nonumber\\
&\times\partial_{k_i}\!\!\left(\frac{1}{i\hbar\omega-\xi_{{\bf k} 2}}\right)
\partial_{k_j}\!\!\left(\frac{1}{i\hbar\omega+\xi_{{\bf k} 2}}\right),
\label{eq.first''A}
\end{align}
with $k_i$ the Cartesian component of ${\bf k}$. As the Q1D dispersion does not depend on $k_y$ and $k_z$, Eq.~(\ref{eq.first''A}) is reduced to 
\begin{align}
{\cal I}_{a2}''= T \frac{\hbar^2}{m_2} 
\partial^2_x\Delta_2\sum_\omega\int \frac{d^3{\bf k}}{(2\pi)^3} \frac{\xi_{{\bf k} 2}+\mu}{\Big(\hbar^2\omega^2+\xi^{2}_{{\bf k} 2}\Big)^2}.
\end{align}
The summation over the Matsubara frequencies yields
\begin{align}
\label{eq.sum2}
\sum_\omega\frac{1}{\left(\hbar^2\omega^2+\xi^{2}_{{\bf k}2}\right)^2}=
&\frac{T \sinh\big(\xi_{\bf k 2}/T\big)-\xi_{{\bf k}2}}{8 \xi^3_{{\bf k}2} T^2} \nonumber\\
&\times\text{sech}^2(\xi_{{\bf k}2}/2 T).
\end{align}
Changing the integration variable to the single-particle energy and keeping only the leading-order term in the $\tau$ expansion, one obtains
\begin{align}
{\cal I}_{a2}'' = {\cal K}^{(x)}_2 \partial^2_x \Delta_2,
\label{eq.first''B}
\end{align}
with
\begin{align}
{\cal K}^{(x)}_2 = &\hbar^2 v_2^2  \frac{N_2}{8\, \hbar^2 \omega_c^2} \int\limits_{-\tilde{\mu}}^{+\infty} d\varepsilon\, \frac{\sqrt{\varepsilon+\tilde{\mu}} }{\varepsilon^3} \, \text{sech}^2  \big(\varepsilon / 2\tilde T_{c0} \big)  \notag \\
& \times \left[ \sinh \left( \frac{\varepsilon}{\tilde T_{c0}} \right)- \frac{\varepsilon}{\tilde T_{c0}}\right]\,\theta(1-|\epsilon|), 
\label{eq.first''C}
\end{align}
which gives ${\cal K}^{(x)}_2$ in Eq.~(\ref{eq.coeff2}) when $\tilde{\mu} \leq 1$. Thus, for the first  term in Eq.~(\ref{eq.gapexp}) we find
\begin{align}
{\cal I}_{a2} =({\cal A}_2 - a_2)\Delta_2 + {\cal K}^{(x)}_2\partial_x^2\Delta_2.   
\label{eq.first-fin}
\end{align}

Finally, we calculate the nonlinear term ${\cal I}_{b2}$ in Eq.~(\ref{eq.gapexp}). It is represented in the form
\begin{align}
{\cal I}_{b2}=&-T\Delta_2 |\Delta_2|^2\sum_\omega\int\frac{\mbox{d}^3k}{(2\pi)^3}\frac{1}{\Big(\hbar^2\omega^2+\xi^2_{{\bf k}2}\Big)^2}.
\end{align}
This expression is evaluated by taking the sum over the Matsubara frequencies, see Eq.~(\ref{eq.Q1D}) and changing the integration variables as previously. Finally, applying the $\tau$ expansion and keeping the leading contribution in $\tau$, we get
\begin{align}
{\cal I}_{b2}=b_2 \Delta_2({\bf r}) |\Delta_2({\bf r})|^2,
\label{eq.second-fin}    
\end{align}
where
\begin{align}
b_2 = & \frac{N_2}{4\hbar^2\omega_c^2} \int\limits_{-\tilde{\mu}}^{+\infty}d\varepsilon\,\frac{\text{sech}^2  \big(\varepsilon/2\tilde T_{c0} \big) }{\varepsilon^3 \sqrt{\varepsilon+\tilde{\mu}}} \nonumber\\
&\times \left[ \sinh \Big( \frac{\varepsilon}{\tilde T_{c0}} \Big)- \frac{\varepsilon}{\tilde T_{c0}}\right] \theta(1-|\varepsilon|).
\end{align}
For $\tilde{\mu} \leq 1$, this expression for $b_2$ coincides with  Eq.~(\ref{eq.coeff2}).

Thus, the anomalous Green function of the Q1D band in the GL approximation is given by
\begin{align}
R_2({\bf r})=&({\cal A}_2 - a_2)\Delta_2 + {\cal K}^{(x)}_2\partial_x^2\Delta_2\nonumber\\
&+ b_2 \Delta_2 |\Delta_2|^2.
\label{eq.R2}
\end{align}
Notice, that the functional derivative of the free energy given by Eq.~(\ref{eq.free1}) yields Eq.~(\ref{eq.selfcons}), where $R_{\nu}$ is given by Eq.~(\ref{eq.R2}).

\end{document}